\documentclass[%
reprint,
superscriptaddress,
amsmath,amssymb,amsthm,mathrsfs,
aps,
prl,
]{revtex4-1}
\usepackage{graphicx}
\usepackage{dcolumn}
\usepackage{bm}
\usepackage[usenames]{color}
\usepackage[ 
margin=0.75in,
]{geometry}
\usepackage{ulem}

\def\BFSe{BaFe$_2$Se$_3$}
\def\BFS{BaFe$_2$S$_3$}
\def\TN{$T_{\textrm{N}}$}
\def\muSR{$\mu$SR}
\def\muB{$\mu_\mathrm{B}$}
\def\Tc{$T_{\textrm{c}}$}

\begin{document}

\title{
Gradual Enhancement of Stripe-Type Antiferromagnetism in Spin Ladder Material BaFe$_2$S$_3$ Under Pressure
}

\author{Liangliang Zheng}
\affiliation{%
	School of Physics, Sun Yat-Sen University, Guangzhou 510275, China.
}%

\author{Benjamin A. Frandsen}
\email{benfrandsen@byu.edu}
\affiliation{%
	Materials Sciences Division, Lawrence Berkeley National Laboratory, Berkeley, California 94720, USA.
}%
	\affiliation{ %
	Department of Physics and Astronomy, Brigham Young University, Provo, Utah 84602, USA.
} %

\author{Changwei Wu}
\affiliation{%
	School of Physics, Sun Yat-Sen University, Guangzhou 510275, China.
}%

\author{Ming Yi}
\affiliation{ %
	Department of Physics, University of California, Berkeley, California 94720, USA.
} %
\affiliation{ %
	Department of Physics and Astronomy, Rice University, Houston, TX 77005, USA
} %

\author{Shan Wu}
\affiliation{ %
	Department of Physics, University of California, Berkeley, California 94720, USA.
} %

\author{Qingzhen Huang}
\affiliation{%
	NIST Center for Neutron Research, National Institute of Standards and Technology, Gaithersburg, MD 208999, USA.
}%

\author{Edith Bourret-Courchesne}
\affiliation{%
	Materials Sciences Division, Lawrence Berkeley National Laboratory, Berkeley, California 94720, USA.
}%

\author{G. Simutis}
\affiliation{%
	Laboratory for Muon Spin Spectroscopy, Paul Scherrer Institut, 5232 Villigen PSI, Switzerland
}%

\author{R. Khasanov}
\affiliation{%
	Laboratory for Muon Spin Spectroscopy, Paul Scherrer Institut, 5232 Villigen PSI, Switzerland
}%

\author{Dao-Xin Yao}
\affiliation{%
	School of Physics, Sun Yat-Sen University, Guangzhou 510275, China.
}%

\author{Meng Wang}
\email{wangmeng5@mail.sysu.edu.cn}
\affiliation{%
	School of Physics, Sun Yat-Sen University, Guangzhou 510275, China.
}%

\author{Robert J. Birgeneau}
\affiliation{ %
	Department of Physics, University of California, Berkeley, California 94720, USA.
} %
\affiliation{%
	Materials Sciences Division, Lawrence Berkeley National Laboratory, Berkeley, California 94720, USA.
}%
\affiliation{ %
	Department of Materials Science and Engineering, University of California, Berkeley, California 94720, USA.
} %

\begin{abstract}
We report pressure-dependent neutron diffraction and muon spin relaxation/rotation measurements combined with first-principles calculations to investigate the structural, magnetic, and electronic properties of BaFe$_2$S$_3$ under pressure. The experimental results reveal a gradual enhancement of the stripe-type ordering temperature with increasing pressure up to 2.6~GPa and no observable change in the size of the ordered moment. The \textit{ab initio} calculations suggest that the magnetism is highly sensitive to the Fe-S bond lengths and angles, clarifying discrepancies with previously published results. In contrast to our experimental observations, the calculations predict a monotonic reduction of the ordered moment with pressure. We suggest that the robustness of the stripe-type antiferromagnetism is due to strong electron correlations not fully considered in the calculations.

\end{abstract}

\maketitle

Iron-based superconductors (FeSCs) remain an outstanding problem in condensed matter physics, eluding a comprehensive explanation despite more than a decade of intensive research. However, there is widespread agreement that the interrelationships among crystal structure, magnetism, and electronic orders of various types are vital for understanding the origin of superconductivity (SC) in these materials~\cite{dagot;rmp13,dai;rmp15,hoson;pc15,si;nrm16,ferna;rpp17}. Until recently, the crystal structures of all known FeSCs were characterized by a two-dimensional (2D) square lattice comprised of edge-sharing Fe$X_4$ tetrahedra ($X$ = Se, P, and As), with SC typically appearing upon carrier doping, isovalent substitution, or application of pressure~\cite{pagli;np10,johns;advp10}. The discoveries of pressure-induced SC in \BFS\ and \BFSe\ have introduced a new, quasi-one-dimensional structural template for FeSCs~\cite{takah;nm15,yamau;prl15,ying;prb17}. In both systems, the Fe$X_4$ edge-sharing tetrahedra create well-separated iron ladders~\cite{hong;jssc72}, distinct from the planar square lattice of other FeSCs. These compounds therefore represent an important new family of materials that has attracted significant theoretical and experimental attention in the quest to understand iron-based superconductivity~\cite{caron;prb11,krzto;jpcm11,caron;prb12,dong;prl14,arita;prb15,hirat;prb15,mouri;prl15,ootsu;prb15,popov;prb15,suzuk;prb15,chi;prl16,patel;prb16,patel;prb17,takub;prb17,wang;prb17ii,zhang;prb17,li;prb18,zhang;prb18}.

\begin{figure}
	\includegraphics[width=85mm]{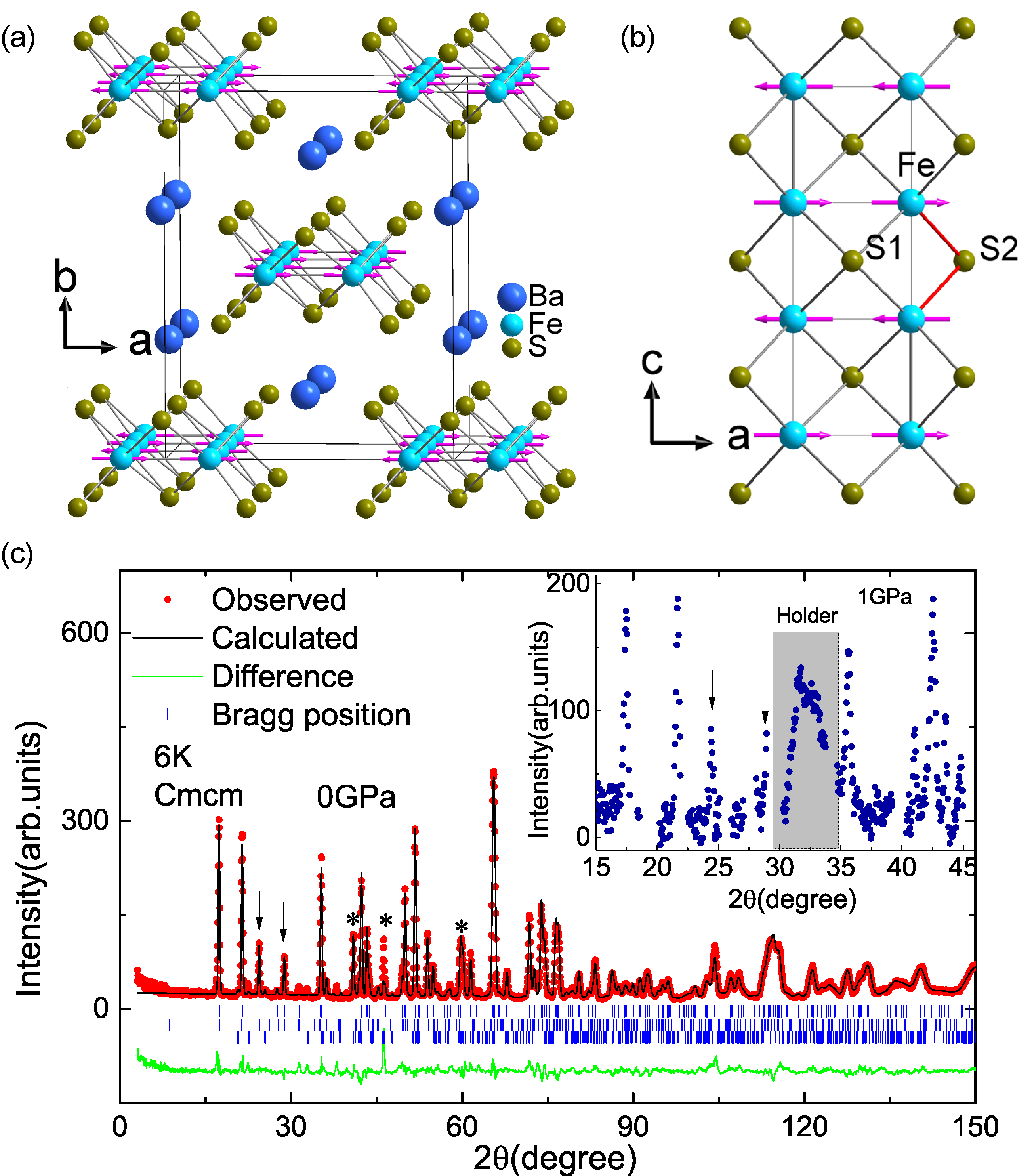}
	\caption{\label{fig:structure}    (a) Crystal and magnetic structure of BaFe$_2$S$_3$ (Ba = dark blue, Fe = light blue, S = green). The solid lines indicate one unit cell. (b) Isolated view of the one-dimensional, edge-sharing FeS$_4$ tetrahedra forming the spin ladders. The Fe$^{2+}$ magnetic moments are aligned ferromagnetically along the rungs and antiferromagnetically along the legs. Two distinct S sites exist, labeled S1 (inside the ladder) and S2 (outside). (c) Neutron powder diffraction pattern for \BFS\ at 6~K and ambient pressure. The pattern was fit with the orthorhombic $Cmcm$ model and the stripe-type magnetic phase. Arrows and asterisks indicate the most prominent peaks from the magnetic order and the $\sim$6\% impurity phase Fe$_7$S$_8$, respectively. From top to bottom, the three rows of tick marks indicate positions of the nuclear, magnetic and impurity Bragg peaks. Inset: Diffractogram collected at 6~K and 1~GPa. The gray rectangle indicates the signal from the pressure cell.}
	
\end{figure}

\BFS\ belongs to the orthorhombic space group $Cmcm$ and orders antiferromagnetically below $\sim$100-120~K in a stripe-type spin configuration~\cite{takah;nm15}. The crystal and magnetic structures are displayed in Fig.~\ref{fig:structure} (a, b).
The antiferromagnetic insulating nature of \BFS\ at ambient pressure gives way to SC under a pressure of $\sim$11~GPa, with a maximum superconducting \Tc\ of $\sim$25~K~\cite{takah;nm15,yamau;prl15}. A recent study of a single-crystal specimen of \BFS\ reported an additional transition at 0.95~GPa, marked by an abrupt increase of the N\'{e}el temperature (\TN) and ordered moment by approximately 50\% and 25\%, respectively, followed by a rapid suppression of \TN\ at higher pressures~\cite{chi;prl16}. This was interpreted as a transition to an orbitally selective Mott phase thought to be the true parent state of the SC at higher pressure. To elucidate this reported transition and its relevance to SC, more studies of \BFS\ under pressure are necessary.
 
In this work, we investigate the effects of pressure on magnetism in \BFS\ using neutron powder diffraction (NPD), muon spin relaxation/rotation (\muSR), and first-principles calculations. The NPD and \muSR\ results show a gradual increase in \TN\ with pressure followed by an eventual plateau around the highest pressure measured (2.6~GPa), with no observable change in the magnitude of the ordered moment. This contrasts with the abrupt magnetic enhancement around 1~GPa and subsequent reduction reported earlier~\cite{chi;prl16}. Our density functional theory (DFT) calculations reveal a delicate sensitivity of the magnetism on the Fe-S bond lengths and angles, which may explain the discrepancy with the earlier work. Finally, the DFT calculations predict a monotonic suppression of the magnetism with pressure, in contrast to the initial enhancement observed experimentally. This may indicate that strong electronic correlations not accounted for in the DFT calculations make the stripe-type antiferromagnetism unexpectedly robust against pressure.

\BFS\ samples were grown by the Bridgman method, forming small, needle-like single crystals a few millimeters in length. The details of the synthesis procedure have been documented previously~\cite{wang;prb14,wang;prb17ii}. Due to the difficulty of co-aligning such crystals, we ground 8~g of single crystals into a powder for subsequent measurement. The NPD experiment was carried out on the High Resolution Powder Diffractometer BT1 at the NIST Center for Neutron Research (NCNR) using a monochromatic beam with $\lambda$ = 2.0779~\AA. A closed cycle refrigerator was used to control the sample temperature. Measurements were performed both at ambient pressure and in a pressure cell with an applied hydrostatic pressure of 1~GPa. Rietveld refinements of the crystal and magnetic structures were conducted using the FullProf Suite~\cite{rodri;pb93}. \muSR\ measurements at ambient pressure were conducted on the LAMPF instrument at TRIUMF in Vancouver, Canada, and the pressure-dependent experiments were performed on the GPD instrument at the Paul Scherrer Institute in Villigen, Switzerland. A gas-flow cryostat was used for temperature control in both cases. Daphne oil was used to transmit hydrostatic pressure, which was calibrated with a superconducting indium plate immersed in the oil with the sample. The uncertainty in the measured pressure was less than 0.1~GPa. The \muSR\ spectra were analyzed using the least-squares minimization routines in the MusrFit software package~\cite{suter;physproc12}. DFT calculations were performed using the Vienna Ab initio Simulation Package (VASP). The electron exchange correlation potential is included in the generalized gradient approximation (GGA) of the Perdew, Burke, and Ernzerhof form~\cite{perde;prl96} through $E_C^{\mathrm{GGA}}[n_{\uparrow},n_{\downarrow}] = \int \mathrm{d}^3rn[\epsilon^{\mathrm{unif}}_C (r_s,\zeta)+H(r_s,\zeta,t)]$, where $r_s$ is the local Seitz radius, $\zeta = (n_{\uparrow}-n_{\downarrow})/n$ is the relative spin polarization, and $t = |\nabla n|/2\phi k_s n$ is the dimensionless density gradient. 

We first present the NPD data. Fig.~\ref{fig:structure}(c) displays the diffraction pattern for \BFS\ at 6~K and ambient pressure. Refinements using the expected orthorhombic structure and stripe-type magnetic order provide a good fit. The refined structural and magnetic parameters at 6~K are summarized in Table 1. Three small Bragg peaks well indexed by Fe$_7$S$_8$ were also observed, indicating an impurity level of about 6\%. Additional unknown impurities may also be present in small amounts ($\lesssim$~1\%), resulting in somewhat less-than-ideal agreement factors.\\

\begin{table}
	\caption{Structural and magnetic parameters of BaFe$_2$S$_3$ at ambient pressure and 6~K. The space group is \textit{Cmcm} (No.63) with $a$=8.7248(2)~\AA, $b$=11.1643(3)~\AA, $c$=5.2618(1)~\AA, and the agreement factors are
		$R_p=15.6\%$, $wR_p=17.5\%$, $\chi^2=4.5\%$.}
	
	\begin{tabular}{ccccccc}
		\hline \hline
		Atom    & Site   & x          & y         & z        & Occ.    &  M($\mu_B$)  \\ \hline
		Ba      & 4c     & 0.50000     & 0.1871(5) & 0.25000   & 1.00000  &       \\
		Fe      & 8e     & 0.3450(2)  & 0.50000 & 0.00000   & 1.053(1)  & 1.29(3) \\
		S1      & 4c     & 0.5000    & 0.609510 & 0.25000   & 1.026(13)  &  \\
		S2      & 8g     & 0.2058(6)    & 0.3750(6) & 0.25000   & 1.005(16)  & \\ \hline \hline
	\end{tabular}
	\label{table:t1}
\end{table}

The refined ordered  moment at 6~K and ambient pressure is 1.29~$\pm$~0.03~\muB\ and lies along the rung direction ($a$ axis), as shown in Fig.~\ref{fig:structure}(a). This is slightly larger than the reported values of 1.02 and 1.20~\muB\ in Refs.~\cite{chi;prl16,takah;nm15}. A portion of the diffraction pattern collected at 6~K and a pressure of 1~GPa is shown in the inset in Fig.~\ref{fig:structure}(c). The data quality is reduced due to the large background from the pressure cell, preventing a satisfactory Rietveld refinement. However, the magnitude of the ordered moment can still be determined through comparison to nuclear Bragg peak intensities, yielding a value of 1.34~$\pm$~0.18~\muB. The magnitude of the ordered magnetic moment at ambient pressure and 1~GPa is thus consistent within error.
\begin{figure}
	\includegraphics[width=80mm]{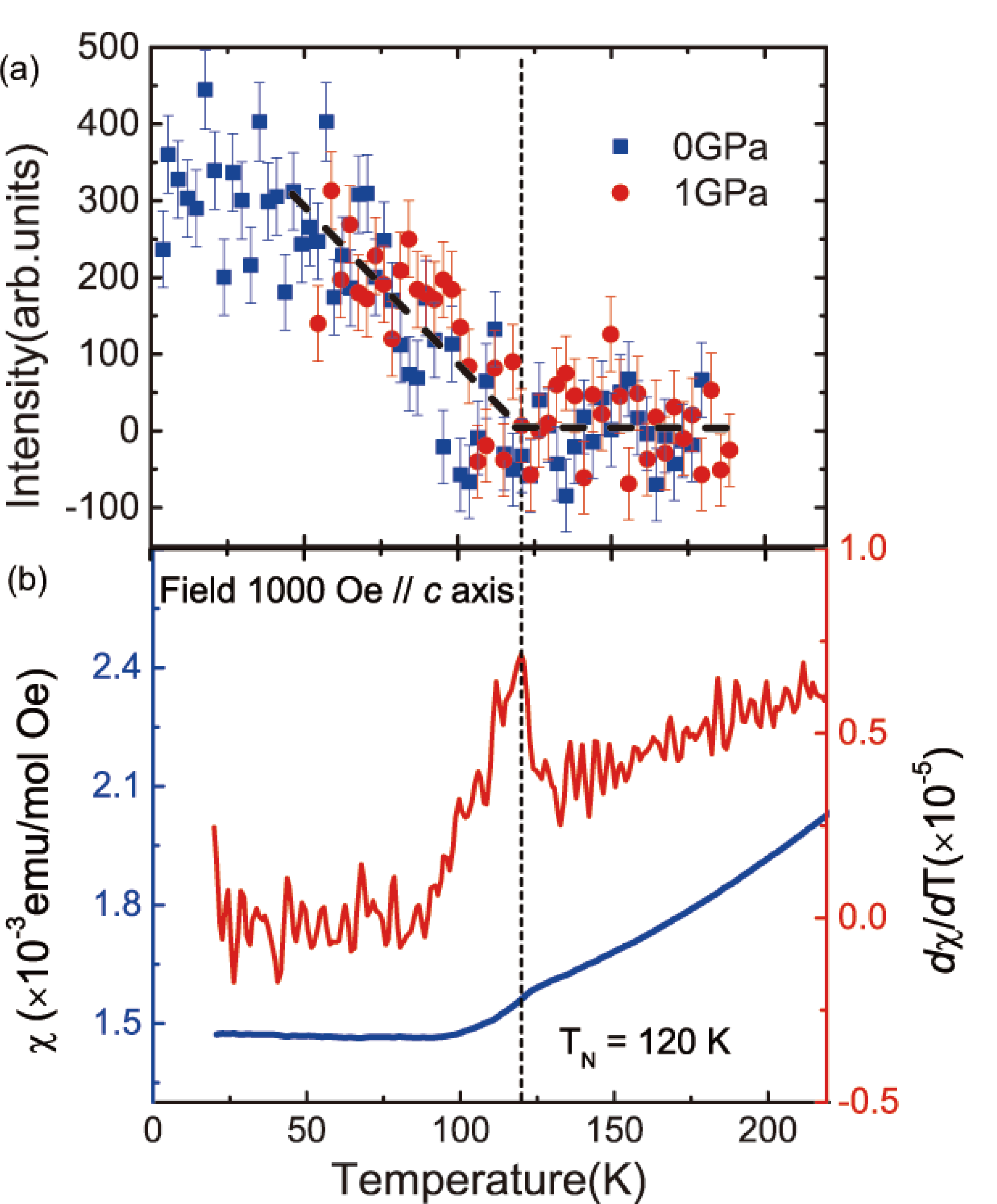}
	\caption{(a) Temperature dependence of the magnetic Bragg peak intensity at 2$\theta$=24.4$^{\circ}$ in ambient pressure and 1~GPa. No significant change is observed in the ordering temperature. (b) DC magnetic susceptibility of \BFS\ (blue curve) and its temperature derivative (red curve). The peak in the derivative indicates a magnetic transition around 100 - 120~K, in agreement with the neutron data.}
	\label{fig:mag}
	
\end{figure}
We next investigated the ordering temperature by measuring the intensity of the magnetic Bragg peak near 2$\theta$~=~24.4$^{\circ}$ as a function of temperature for both ambient pressure and 1~GPa, shown in Fig.~\ref{fig:mag}(a). The intensity begins to increase below \TN~$\approx$~100 - 120~K for both pressures. The dashed line is a guide to the eye, showing an upturn at 120~K. The DC zero-field-cooled magnetic susceptibility measured at ambient pressure with $H = 1000$Oe, shown along with its temperature derivative in Fig.~\ref{fig:mag}(b), exhibits a broad feature confirming \TN~$\approx$~100 - 120~K. This ordering temperature is generally consistent with the earlier reports of \TN~$=$~105~K \cite{chi;prl16} and 119~K \cite{takah;nm15}.

\begin{figure}
	\includegraphics[width=90mm]{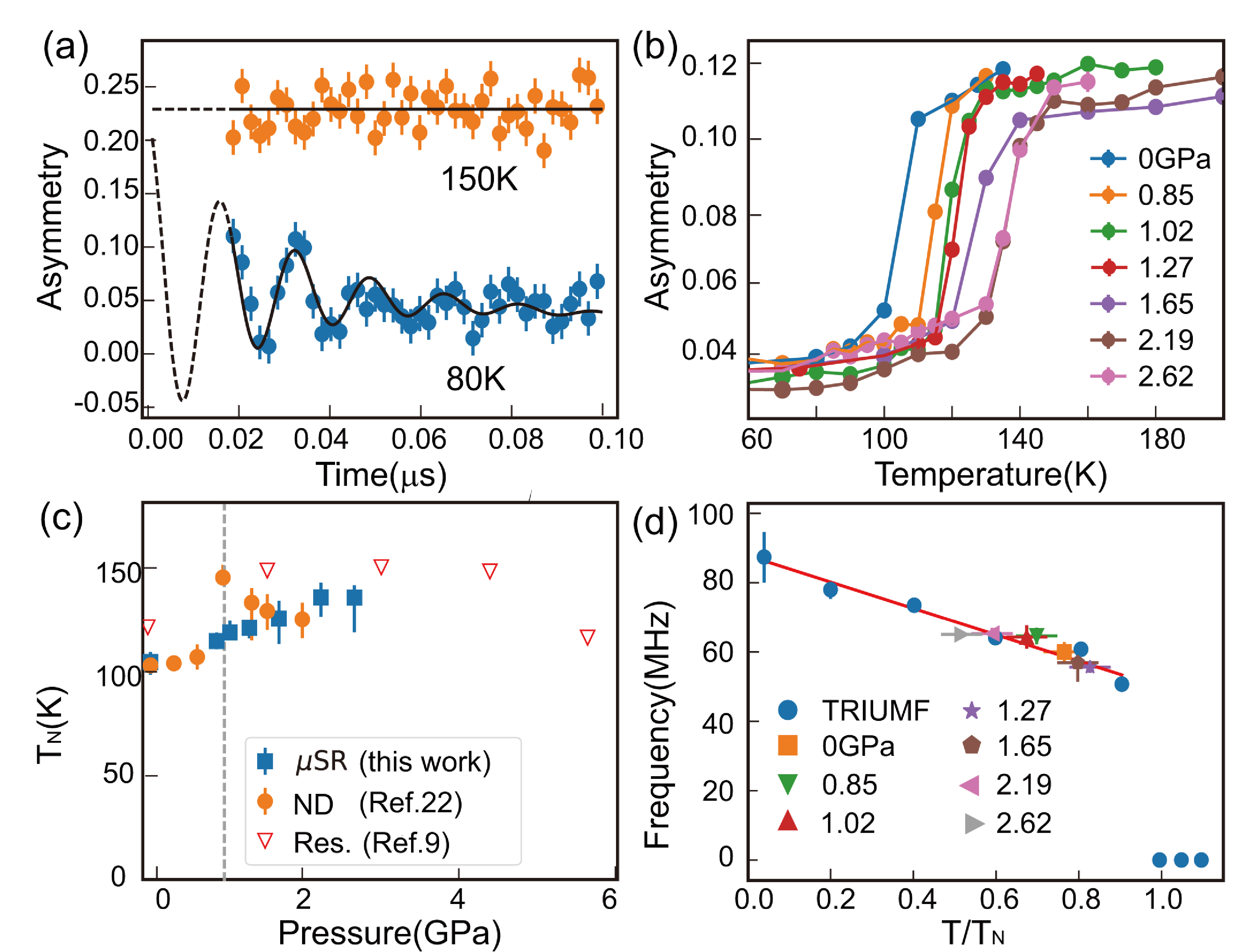}
	\caption{\label{fig:muSR}  (a) Representative zero-field \muSR\ spectra for \BFS\ at 150~K (above \TN) and 80~K (below \TN). Filled circles represent data, black curves represent the fits. The dashed black curves represent extrapolations of the fits from the start of the data window at $t$ = 0.02 $\mu$s to $t$ = 0. (b) Amplitude of the non-oscillating asymmetry component of the ZF \muSR\ spectra as a function of temperature for various applied pressures. (c) \TN\ as a function of pressure obtained from the present \muSR\ measurements (blue squares) and previously reported neutron diffraction (orange circles; Ref.~\cite{chi;prl16}) and resistivity measurements (red triangles; Ref.~\cite{yamau;prl15}). The dashed vertical line represents the critical pressure for the abrupt transition reported in Ref.~\cite{chi;prl16}. ND = neutron diffraction. (d) Refined ZF oscillation frequencies for the data collected at ambient pressure at TRIUMF (blue circles) and at different applied pressures at PSI (various symbols), plotted versus $T/T_{\mathrm{N}}$ for meaningful comparison.}
	
\end{figure}

The NPD results reported so far provide no indication that a change to the ordered moment or ordering temperature occurs between ambient pressure and 1~GPa, in contrast to the dramatic enhancement at 0.95~GPa reported previously~\cite{chi;prl16}. To verify our NPD findings, we now turn to the \muSR\ measurements, which we performed on a sample taken from the same powder batch as was used for the NPD experiments. In Fig.~\ref{fig:muSR}(a), we display two representative time spectra measured in zero external field (ZF) at 150 K (orange) and 80 K (blue), which are respectively above and below the AF transition.
These spectra were collected in the low-background sample environment at TRIUMF. Oscillations of the asymmetry at 80~K originate from the long-range magnetic order, while the featureless spectrum at 150~K reflects the paramagnetic state. The \muSR\ oscillation frequency as a function of dimensionless temperature $T$/\TN\ is given by the blue circles in Fig.~\ref{fig:muSR}(d), showing a steady increase as the temperature decreases and the ordered moment grows.

In Fig.~\ref{fig:muSR}(b), we plot the temperature dependence of the non-oscillating asymmetry component extracted from the ZF spectra at various applied pressures up to 2.6~GPa. The non-oscillating asymmetry is proportional to the paramagnetic volume fraction in the sample, so a bulk magnetic transition is manifest as a significant drop in this asymmetry component as the temperature is lowered. The general trend is a gradual increase in the magnetic transition temperature as the applied pressure is raised from 0 to 2.6~GPa. Interpolating between the data points at each pressure and choosing \TN\ as the temperature where 50\% of the sample is magnetically ordered allows us to quantify \TN\ as a function of pressure, which we illustrate with the blue squares in Fig.~\ref{fig:muSR}(c). At ambient pressure, \TN\ determined from the \muSR\ data in this way is 105~K, consistent with the susceptibility and NPD results. The upper and lower ends of the error bars in Fig.~\ref{fig:muSR}(c) correspond to the temperatures where 20\% and 80\% of the asymmetry has been lost, respectively, giving a good indication of the temperature range of the bulk transition. 

The \muSR\ results clearly show a smooth increase of \TN\ with pressure, possibly plateauing around 2 GPa. The increase in \TN\ between ambient pressure and 1~GPa is not large ($\lesssim 10$ K) and the transition is somewhat broad, perhaps explaining why it was not observed by our neutron measurements. Overall, the pressure dependence of \TN\ determined by \muSR\ is in good qualitative agreement with the kink in resistivity measurements under pressure reported in Ref.~\cite{yamau;prl15}, shown as open red triangles in Fig.~\ref{fig:muSR}(c). Differing sample synthesis methods are known to result in different ordering temperatures, likely explaining the $\sim$20-K offset between our \muSR\ measurements and the resistivity measurements from Ref.~\cite{yamau;prl15}. These results contrast starkly with the earlier report~\cite{chi;prl16} of the dramatic enhancement at 0.95~GPa (vertical dashed line) and subsequent suppression at higher pressure, shown as orange circles in Fig.~\ref{fig:muSR}(c). In Fig.~\ref{fig:muSR}(d), we plot the refined ZF oscillation frequencies determined at each pressure. Due to time constraints, high-quality data sufficient for resolution of the oscillations under pressure were obtained only at one temperature point per applied pressure. To make a meaningful comparison among different pressures and the more complete set of oscillation frequencies measured at TRIUMF, the temperatures have been normalized by \TN. Displayed this way, the frequencies all lie along the same line. Since the oscillation frequency is proportional to the ordered moment, these results indicate there is no dramatic pressure dependence of the moment size.\\

The NPD and \muSR\ results are therefore consistent, demonstrating that there is only a slow and gradual increase of \TN\ with pressure up to $\sim$2~GPa, without any change in the ordered moment. Compared to the results reported in Ref.~\cite{chi;prl16}, our sample exhibits a larger ordered moment at ambient pressure (1.3~\muB\ versus 1.0~\muB) and a markedly different response to pressure (gradual enhancement, not dramatic change). We now explore possible explanations for these discrepancies. First, it is known that differing synthetic procedures can influence the precise stoichiometry, which in turn affects the electronic and magnetic properties~\cite{takah;nm15,hirat;prb15}. In addition, discrepancies in the unit cell have been reported in the literature. Specifically, the $a$ and $b$ lattice parameters in Ref.~\cite{chi;prl16} are $\sim$0.1~\AA\ longer than those in the present work, causing the bond lengths and angles between the Fe and S2 atoms to show obvious differences between the two samples. The Fe-S2 bond length in the current work is nearly 0.09~\AA\ shorter than previously reported (2.2700 versus 2.3587~\AA), and the present Fe-S2-Fe angle is nearly 3$^{\circ}$ larger (70.829$^{\circ}$ versus 67.985$^{\circ}$). The Fe-S2 geometry may therefore influence the magnetism, likely by affecting the magnetic exchange interaction and orbital hybridization between Fe and S atoms. This is supported by our DFT calculations, which show an enhancement of the ordered moment when the Fe-S2 bond length shortens, consistent with the experimental observation.

\begin{figure}
	\includegraphics[width=85mm]{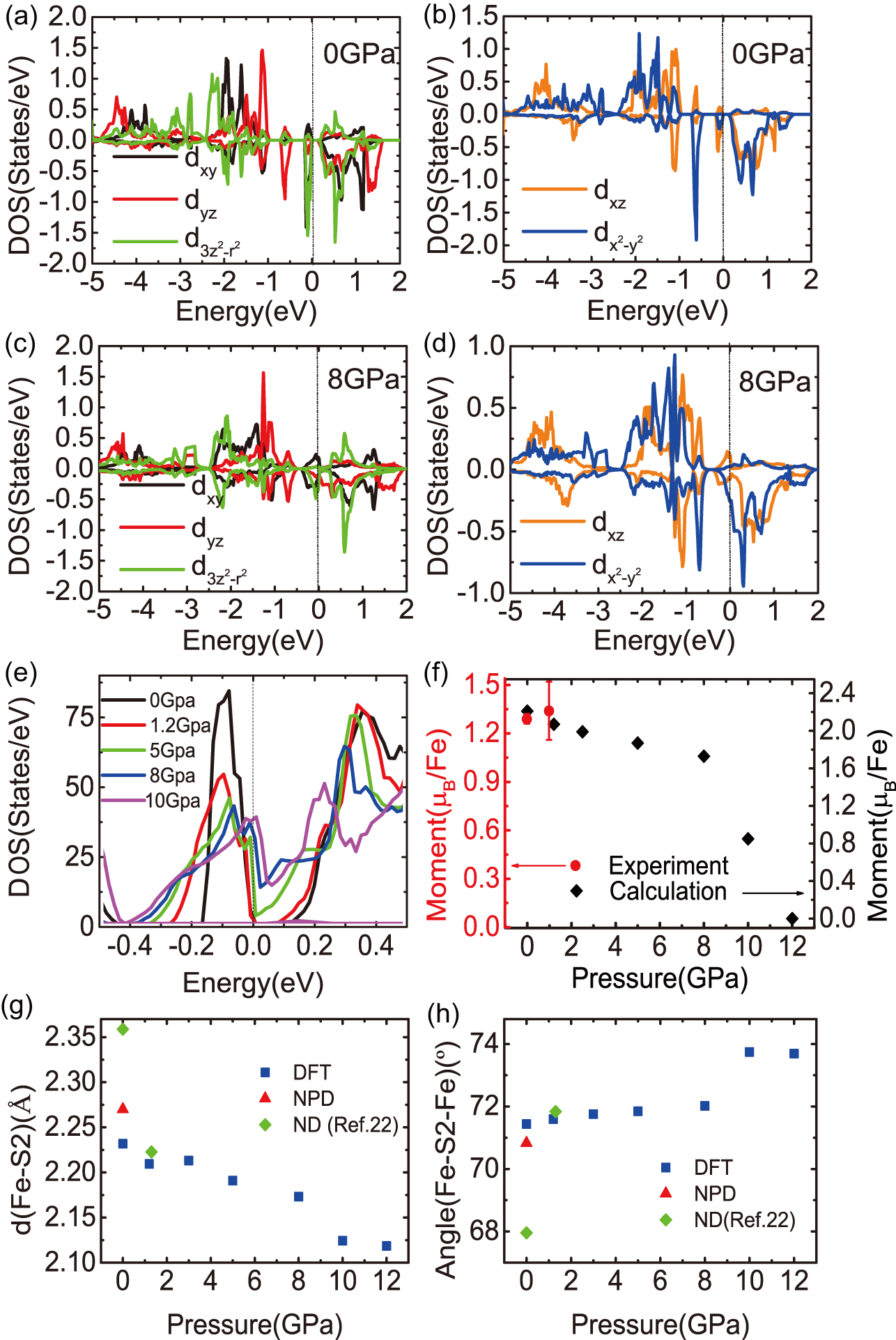}
	\caption{\label{fig:DFT}  (a)-(d) DFT calculations of the density of states (DOS) of Fe 3$d$ orbitals for $P$ = 0~GPa and 8~GPa. (e) Pressure dependence of the total DOS. (f) Calculated (right axis) and observed (left axis) ordered magnetic moments as a function of pressure. Note the difference in scale on the vertical axes. (g) The Fe-S2 bond length as a function of pressure calculated by DFT and compared to experimentally determined values from this work and Ref.~\onlinecite{chi;prl16}. NPD = neutron powder diffraction, ND = neutron diffraction. (h) Same as (g), but displaying the Fe-S2-Fe bond angle.}
	
\end{figure}

We now report our DFT calculations investigating the magnetism under pressure. In Fig.~\ref{fig:DFT}(a)-(d), we display the orbitally resolved density of states (DOS) of the Fe 3$d$ orbitals for zero pressure and 8~GPa. For these calculations, the starting lattice parameters at zero pressure were set to the experimentally determined values.
The various colors indicate different Fe 3$d$ orbitals and the dashed lines represent the Fermi energy. The calculations correctly predict zero DOS at the Fermi energy at 0~GPa, corresponding to an insulating state, whereas metallic behavior with a finite DOS at the Fermi energy is expected for 8~GPa. We found a large difference in the DOS contributed by up and down spins in each orbital, implying that all of the Fe 3$d$ orbitals contribute to the magnetism. The calculated contributions of each orbital to the spin moment of each Fe atom per formula unit are 0.40 ($d_{xy}$), 0.44 ($d_{yz}$), 0.36 ($d_{3z^2-r^2}$), 0.52 ($d_{xz}$), and 0.51 ($d_{x^2-y^2}$) at 0~GPa and 0.30 ($d_{xy}$), 0.32 ($d_{yz}$), 0.28 ($d_{3z^2-r^2}$), 0.38 ($d_{xz}$), and 0.42 ($d_{x^2-y^2}$) at 8~GPa. We see that $d_{x^2-y^2}$ and $d_{xz}$ contribute the most to the calculated spin moment, and the calculated contribution from all orbitals decreases as pressure increases. Fig.~\ref{fig:DFT}(e) shows the total DOS under different pressures for the stripe AF phase, showing an insulator-metal transition between 1.2 and 5~GPa. Additional calculations on a finer pressure grid indicate the insulating gap closes at 4.0~GPa, marking the metal-insulator transition as predicted by DFT.

DFT calculations performed for five distinct ferromagnetic and antiferromagnetic orders correctly predicted the stripe-type antiferromagnetic structure to be most stable. The calculated ordered moment on the iron atoms at different pressures is shown as black diamonds (right axis) in Fig.~\ref{fig:DFT}(f), together with the measured ordered moment from the neutron experiments at ambient pressure and 1~GPa shown as red circles (left axis). Note the difference in scale on the two vertical axes. The calculations predict a monotonic decrease of the ordered moment with pressure, consistent with earlier first-principles studies~\cite{suzuk;prb15,zhang;prb17}, with a change of slope between 8 and 10~GPa and a complete suppression of the ordered moment at 12~GPa. This is near the pressure at which SC appears. The calculated magnitude of the ordered moment at 0~GPa is nearly a factor of 2 larger than the experimental result, which likely results largely from strong zero-point fluctuations in this quasi-one-dimensional system. The significant reduction of \TN\ from the value predicted by DFT (1800~K) further supports this possibility. Another contributing factor to the reduced moment may be the coexistence of localized iron spins and itinerant electrons~\cite{ootsu;prb15}. Additionally, the stability of the ordered moment revealed by NPD and \muSR\ experiments up to 2.6~GPa may indicate that stripe-type magnetism is more robust in the low-pressure regime than expected on the basis of DFT calculations. Given that moderately strong electron correlations are expected to exist in \BFS~\cite{arita;prb15}, we suggest correlation effects not properly accounted for in the DFT calculations may be responsible for this. Finally, Fig.~\ref{fig:DFT}(g) and (h) display pressure-dependent DFT calculations of the Fe-S2 bond length and Fe-S2-Fe bond angle, respectively, along with the experimentally observed values in this work and Ref.~\onlinecite{chi;prl16}. The anomalous bond geometry at 0~GPa of the sample in Ref.~\onlinecite{chi;prl16} is clearly illustrated. Interestingly, the DFT calculations show a step change in the bond geometry between 8 and 10~GPa, coincident with the enhanced suppression of the magnetic moment evident in Fig.~\ref{fig:DFT}(f). This reflects the strong magnetoelastic coupling inherent in this system.

In summary, we have studied the magnetic and electronic properties of \BFS\ through neutron powder diffraction, \muSR, and DFT calculations. Our results confirm the stripe-type antiferromagnetic structure, but with a slightly larger ordered moment than reported in Ref.~\cite{chi;prl16}. This appears to be due to a difference between the bond lengths and angles of Fe and S2. The experimental results show a gradual increase in \TN\ up to 2.6~GPa with no change in the ordered moment, in contrast to the dramatic increase in \TN\ and the ordered moment reported previously at 0.95~GPa. Given this lack of consistency, interpretations of the previous results in terms of orbitally selective Mott physics should be taken with caution. Our DFT calculations correctly yield an insulating ground state at ambient pressure, with an insulator-metal transition predicted around 4~GPa. The calculations predict the ordered moment to decrease to zero around 12~GPa, while the observed robustness of the moment and ordering temperature at pressures up to 2.6~GPa suggests that electron correlation effects fortify the stripe-type antiferromagnetism.

\begin{acknowledgments}
\textbf{Acknowledgements}

Work at Lawrence Berkeley National Laboratory was funded by the U.S. Department of Energy, Office of Science, Office of Basic Energy Sciences, Materials Sciences and Engineering Division under Contract No. DE-AC02-05-CH11231 within the Quantum Materials Program (KC2202).  M.W. and L.Z. were supported by the Hundreds of Talents program of Sun Yat-Sen University, National Science Fund of Guangdong under Contract No. 2018A030313055, and Young Zhujiang Scholar program. D.X.Y. and C.W. acknowledge support from NKRDPC-3272017YFA0206203, NKRDPC-2018YFA0306001, NSFC-11574404, National Supercomputer Center in Guangzhou, and Leading Talent Program of Guangdong Special Projects. The work of G.S. was supported by the Swiss National Science Foundation grants 200021-149486 and 200021-175935. Part of this work is based on experiments performed at the Swiss Muon Source S$\mu$S, Paul Scherrer Institute, Villigen, Switzerland, and at the Centre for Material and Molecular Science at TRIUMF in Vancouver, Canada. 
\end{acknowledgments}

\end{document}